# Generation of picosecond pulsed coherent state superpositions

**Ruifang Dong, Anders Tipsmark, Amine Laghaout, Leonid A. Krivitsky, Miroslav Ježek and Ulrik L. Andersen**

*Department of Physics, Technical University of Denmark, Fysikvej, 2800 Kgs. Lyngby, Denmark*

*Quantum Frequency Standards Division, National Time Service Center (NTSC), Chinese Academy of Sciences, 710600 Lintong, Shaanxi, China*

*Data Storage Institute, Agency for Science Technology and Research (A\*STAR), 117608 Singapore*

*Department of Optics, Palacký University, 17. listopadu 12, 77146 Olomouc, Czech Republic*

**Abstract:** We present the generation of approximated coherent state superpositions - referred to as Schrödinger cat states - by the process of subtracting single photons from picosecond pulsed squeezed states of light at 830 nm. The squeezed vacuum states are produced by spontaneous parametric down-conversion (SPDC) in a periodically poled $KTiOPO_4$ crystal while the single photons are probabilistically subtracted using a beamsplitter and a single photon detector. The resulting states are fully characterized with time-resolved homodyne quantum state tomography. Varying the pump power of the SPDC, we generated different states which exhibit non-Gaussian behavior.

## References and links

1. L. A. Wu, H. J. Kimble, J. Hall and Huifa Wu, "Generation of Squeezed States by Parametric Down Conversion", Phys. Rev. Lett. **57**, 2520–2523(1986), http://link.aps.org/doi/10.1103/PhysRevLett.57.2520.
2. H. Vahlbruch, M. Mehmet, S. Chelkowski, B. Hage, A. Franzen, N. Lastzka, S. Goßler, K. Danzmann and R. Schnabel, "Observation of Squeezed Light with 10-dB Quantum-Noise Reduction", Phys. Rev. Lett. **100**, 033602(2008), http://link.aps.org/doi/10.1103/PhysRevLett.100.033602.
3. Y. Takeno, M. Yukawa, H. Yonezawa, and A. Furusawa. "Observation of -9 dB quadrature squeezing with improvement of phase stability in homodyne measurement", Opt. Express **15**, 4321(2007), http://www.opticsinfobase.org/abstract.cfm?URI=oe-15-7-4321.
4. R. Dong, J. Heersink, J. F. Corney, P. D. Drummond, U. L. Andersen and G. Leuchs, "Experimental evidence for Raman-induced limits to efficient squeezing in optical fibers", Opt. Lett. **33**, 116–118(2008), http://www.opticsinfobase.org/abstract.cfm?URI=ol-33-2-116.
5. M. G. A. Paris, "Displacement operator by beam splitter", Phys. Lett. **217**, 78–80(1996), http://linkinghub.elsevier.com/retrieve/pii/0375960196003398.
6. S. Lloyd, "Coherent quantum feedback", Phys. Rev. A **62**, 022108(2000), http://link.aps.org/doi/10.1103/PhysRevA.62.022108.
7. S. Scheel, W. Munro, J. Eisert, K. Nemoto and P. Kok, "Feed-forward and its role in conditional linear optical quantum dynamics", Phys. Rev. A **73**, 034301(2006), http://link.aps.org/doi/10.1103/PhysRevA.73.034301.
8. H. M. Wiseman and G. J. Milburn, "Quantum theory of optical feedback via homodyne detection", Phys. Rev. Lett. **70**, 548–551(1993), http://link.aps.org/doi/10.1103/PhysRevLett.70.548.

9. A. Furusawa, S. L. Braunstein, J. L. Sørensen, C. A. Fuchs, H. J. Kimble and E. S. Polzik, "Unconditional Quantum Teleportation", Science **282**, 706–709(1998), http://www.sciencemag.org/cgi/doi/10.1126/science.282.5389.706.
10. F. Grosshans, G. V. Assche, J. Wenger, R. Brouri, N. J. Cerf and P. Grangier, "Quantum key distribution using gaussian-modulated coherent states", Nature (London) **421**, 238–241(2003),http://www.nature.com/nature/journal/v421/n6920/full/nature01289.html.
11. U. L. Andersen, V. Josse, and G. Leuchs, "Unconditional Quantum Cloning of Coherent States with Linear Optics", Phys. Rev. Lett. **94**, 240503(2005), http://link.aps.org/doi/10.1103/PhysRevLett.94.240503.
12. A. M. Lance, T. Symul, W. P. Bowen, B. Sanders, T. Tyc, T. C. Ralph, and P. K. Lam, "Continuous-variable quantum-state sharing via quantum disentanglement", Phys. Rev. Lett. **71**, 033814(2005), http://link.aps.org/doi/10.1103/PhysRevLett.71.033814.
13. R. Ukai, N. Iwata, Y. Shimokawa, S. Armstrong, A. Politi, J. Yoshikawa, P. V. Loock, and A. Furusawa, "Demonstration of Unconditional One-Way Quantum Computations for Continuous Variables", Phys. Rev. Lett. **103**, 240504(2011), http://link.aps.org/doi/10.1103/PhysRevLett.106.240504.
14. C. Weedbrook, S. Pirandola, R. García-Patrón, N. J. Cerf, T. C. Ralph, J. H. Shapiro, and S. Lloyd, "Gaussian Quantum Information", Rev. Mod. Phys. **84**, 621(2012), http://link.aps.org/doi/10.1103/RevModPhys.84.621.
15. J. Eisert, S. Scheel, M. Plenio, "Distilling Gaussian States with Gaussian Operations is Impossible", Phys. Rev. Lett. **89**, 137903(2002), http://link.aps.org/doi/10.1103/PhysRevLett.89.137903.
16. J. Fiurášek, "Gaussian Transformations and Distillation of Entangled Gaussian States", Phys. Rev. Lett. **89**, 137904(2002), http://link.aps.org/doi/10.1103/PhysRevLett.89.137904.
17. G. Giedke, and J. I. Cirac, "Characterization of Gaussian operations and distillation of Gaussian states", Phys. Rev. A **66**, 032316(2002), http://link.aps.org/doi/10.1103/PhysRevA.66.032316.
18. J. Niset, J. Fiurášek, and N. J. Cerf, "No-Go Theorem for Gaussian Quantum Error Correction", Phys. Rev. Lett. **102**, 120501(2009), http://link.aps.org/doi/10.1103/PhysRevLett.102.120501.
19. S. Lloyd, and S. L. Braunstein, "Quantum Computation over Continuous Variables", Phys. Rev. Lett. **82**, 1784–1787(1999), http://link.aps.org/doi/10.1103/PhysRevLett.82.1784.
20. M.Ohliger, K. Kieling, and J. Eisert, "Limitations of quantum computing with Gaussian cluster states", Phys. Rev. A **82**, 042336(2010), http://link.aps.org/doi/10.1103/PhysRevA.82.042336.
21. L. Magnin, F. Magniez, A. Leverrier, N. J. Cerf, Phys. Rev. A **81**, 010302(2010), http://link.aps.org/doi/10.1103/PhysRevA.81.010302.
22. J. Bell, *Speakable and Unspeakable in Quantum Mechanics* (Cambridge University Press 1987).
23. N. Menicucci, P. van Loock, M. Gu, C. Weedbrook, T. C. Ralph, and M. A. Nielsen, "Universal Quantum Computation with Continuous-Variable Cluster States", Phys. Rev. Lett. **97**, 110501(2006), http://link.aps.org/doi/10.1103/PhysRevLett.97.110501.
24. A. Ourjoumtsev, R. Tualle-brouri, P. Grangier, Philippe and A. Dantan, "Increasing entanglement between Gaussian states by coherent photon subtraction", Phys. Rev. Lett. **98**, 030502(2007), http://link.aps.org/doi/10.1103/PhysRevLett.98.030502.
25. H. Takahashi, J. S. Neergaard-Nielsen, M. Takeuchi, M. Takeoka, K. Hayasaka, A. Furusawa, and M. Sasaki, "Entanglement distillation from Gaussian input states", Nature Photonics **4**, 178–181(2010), http://dx.doi.org/10.1038/nphoton.2010.1.
26. M. Lassen, M. Sabuncu, A. Huck, J. Niset, G. Leuchs, N. J. Cerf, and U. L. Andersen, "Quantum optical coherence can survive photon losses using a continuous-variable quantum erasure-correcting code", Nature Photonics **4**, 700–705(2010), http://dx.doi.org/10.1038/nphoton.2010.168.
27. T.C. Ralph, A. Gilchrist, G.J. Milburn, W.J. Munro, and S. Glancy, "Quantum computation with optical coherent states", Phys. Rev. A **68**, 042319(2003), http://link.aps.org/doi/10.1103/PhysRevA.68.042319.
28. A. P. Lund, H. Jeong, T. C. Ralph, and M. S. Kim, "Conditional production of superpositions of coherent states with inefficient photon detection", Phys. Rev. A **70**, 020101(2004), http://link.aps.org/doi/10.1103/PhysRevA.70.020101.
29. A. Gilchrist, K. Nemoto, W. Munro, T. C. Ralph, S. Glancy, S. L. Braunstein, G. J. and Milburn, "Schrödinger cats and their power for quantum information processing", J. Opt. B: Quantum Semiclass. Opt. **6**,828–833(2004), http://iopscience.iop.org/1464-4266/6/8/032/.
30. T. C. Ralph, "Quantum error correction of continuous-variable states against Gaussian noise", Phys. Rev. A **84**, 022309(2011), http://link.aps.org/doi/10.1103/PhysRevA.84.022339.
31. R. García-Patrón, J. Fiurášek, N. J. Cerf, J. Wenger, R. Tualle-Brouri, and P. Grangier, "Proposal

for a Loophole-Free Bell Test Using Homodyne Detection", Phys. Rev. Lett. **93**, 130409(2004), http://link.aps.org/doi/10.1103/PhysRevLett.93.130409.
32. A. Mandilara, and N. J. Cerf, "Quantum bit commitment under Gaussian constraints", Phys. Rev. A **85**, 062310(2012), http://link.aps.org/doi/10.1103/PhysRevA.85.062310.
33. A. I. Lvovsky, H. Hansen, T. Aichele, O. Benson, J. Mlynek, and S. Schiller, "Quantum State Reconstruction of the Single-Photon Fock State", Phys. Rev. Lett. **87**, 050402(2001), http://link.aps.org/doi/10.1103/PhysRevLett.87.050402.
34. T. Aichele, A. I. Lvovsky and S. Schiller, "Optical mode characterization of single photons prepared by means of conditional measurements on a biphoton state", European Physical Journal D **18**, 237–245(2002), http://link.springer.com/article/10.1140/epjd/e20020028.
35. A. Ourjoumtsev, R. Tualle-Brouri, and P. Grangier, "Quantum Homodyne Tomography of a Two-Photon Fock State", Phys. Rev. Lett. **96**, 213601(2006), http://link.aps.org/doi/10.1103/PhysRevLett.96.213601.
36. S. R. Huisman, N. Jain, Nitin, S. A. Babichev, F. Vewinger, A. N. Zhang, S. H. Youn, and A. I. Lvovsky, "Instant single-photon Fock state tomography", Opt. Lett. **34**, 2739–2741(2009), http://www.opticsinfobase.org/abstract.cfm?URI=ol-34-18-2739
37. , M. Cooper, L. J. Wright, C. Söller, and B. J. Smith, "Experimental generation of multi-photon Fock states", Opt. Express **21**, 5309–53172013, http://dx.doi.org/10.1364/OE.21.005309.
38. K. Laiho1, K. N. Cassemiro, D. Gross, and C. Silberhorn, "Probing the Negative Wigner Function of a Pulsed Single Photon Point by Point", Phys. Rev. Lett. **105**, 253603(2010), http://link.aps.org/doi/10.1103/PhysRevLett.105.253603.
39. P. Marek, J. Fiurášek, "Elementary gates for quantum information with superposed coherent states", Phys. Rev. A **82**, 014304(2010), http://link.aps.org/doi/10.1103/PhysRevA.82.014304.
40. A. Tipsmark, R. Dong, A. Laghaout, P. Marek, M. Ježek, and U. L. Andersen, "Experimental demonstration of a Hadamard gate for coherent state qubits", Phys. Rev. A **84**, 050301(2011), http://link.aps.org/doi/10.1103/PhysRevA.84.050301.
41. R. Blandino, F. Ferreyrol, M. Barbieri, P. Grangier, and R. Tualle-Brouri, "Characterization of a ęÐ-phase shift quantum gate for coherent-state qubits", New J. Physics **14**, 013017(2012), http://iopscience.iop.org/1367-2630/14/1/013017/article.
42. M. Dakna, T. Anhut, Opatrný, Knöll, and D. G. Welsch, "Generating Schrödinger-cat-like states by means of conditional measurements on a beam splitter", Phys. Rev. A **55**, 3184–3194(1997), http://link.aps.org/doi/10.1103/PhysRevA.55.3184.
43. H. Takahashi, K. Wakui, S. Suzuki, M. Takeoka, K. Hayasaka, A. Furusawa, M. Sasaki, "Generation of Large-Amplitude Coherent-State Superposition via Ancilla-Assisted Photon Subtraction", Phys. Rev. Lett. **101**, 233605(2008), http://link.aps.org/doi/10.1103/PhysRevLett.101.233605.
44. J. B. Brask, R. Chaves, and N. Brunner, "Testing nonlocality of a single photon without a shared reference frame", Phys. Rev. A **88**, 012111(2013), http://link.aps.org/doi/10.1103/PhysRevA.88.012111.
45. A. Laghaout, J. S. Neergaard-Nielsen, I. Rigas, C. Kragh, A. Tipsmark, and U. L. Andersen, "Amplification of realistic Schrödinger-cat-state-like states by homodyne heralding", Phys. Rev. A **87**, 043826(2013), http://link.aps.org/doi/10.1103/PhysRevA.87.043826.
46. A. Ourjoumtsev, H. Jeong, R. Tualle-brouri, and P. Grangier, "Generation of Optical "Schrödinger Cats" from Photon Number Stataes", Nature **448**, 784–786(2007), http://www.nature.com/nature/journal/v448/n7155/full/nature06054.html
47. J. Wenger, R. Tualle-Brouri, and P. Grangier. "Pulsed homodyne measurements of femtosecond squeezed pulses generated by single-pass parametric deamplification", Opt. Lett. **29**, 1267–1269(2004), http://www.opticsinfobase.org/abstract.cfm?URI=OL-29-11-1267.
48. J. S. Neergaard-Nielsen, B. Melholt Nielsen, C. Hettich, K. Mølmer, and E. S. Polzik. "Generation of a Superposition of Odd Photon Number States for Quantum Information Networks", Phys. Rev. Lett. **97**, 083604(2006), http://link.aps.org/doi/10.1103/PhysRevLett.97.083604.
49. K. Wakui, H. Takahashi, A. Furusawa, and M. Sasaki. "Photon subtracted squeezed states generated with periodically poled $KTiOPO_4$", Opt. Express **15**, 3568–3574(2007), http://www.opticsinfobase.org/abstract.cfm?URI=oe-15-6-3568.
50. T. Gerrits, S. Glancy, T. S. Clement, B. Calkins, A. E. Lita, A. J. Miller, A. L. Migdall, S. W. Nam, R. P. Mirin, and E. Knill1, "Generation of Optical Coherent State Superpositions by Number-Resolved Photon Subtraction from Squeezed Vacuum", Phys. Rev. A **82**, 031802(2010), http://link.aps.org/doi/10.1103/PhysRevA.82.031802.
51. W. P. Bowen, R. Schnabel, H. A. Bachor, and P. K. Lam, "Non-Gaussian operation based on photon subtraction using a photon-number-resolving detector at a telecommunications wavelength", Nature Photonics **4**, 655–660(2010),

http://www.nature.com/doifinder/10.1038/nphoton.2010.158.
52. J. S. Neergaard-Nielsen, M. Takeuchi, K. Wakui, H. Takahashi, K. Hayasaka, M. Takeoka, and M. Sasaki, "Optical Continuous-Variable Qubit", Phys. Rev. Lett. **105**, 053602(2010), http://link.aps.org/doi/10.1103/PhysRevLett.105.053602
53. N. Lee, H. Benichi, Y. Takeno, S. Takeda, J. Webb, E. Huntington, A. Furusawa, "Teleportation of Nonclassical Wave Packets of Light", Science **332**, 330 (2011), http://www.sciencemag.org/cgi/doi/10.1126/science.1201034.
54. G. D. Boyd, and D. A. Kleinman, "Parametric Interaction of Focused Gaussian Light Beams", Journal of Applied Physics **39**, 3597(1968), http://link.aip.org/link/?JAP/39/3597/1&Agg=doi
55. J. Wenger, R. Tualle-Brouri, and P. Grangier, "Pulsed homodyne measurements of femtosecond squeezed pulses generated by single-pass parametric deamplification", Opt. Lett. **29**, 1267–1269(2004), http://www.opticsinfobase.org/abstract.cfm?URI=OL-29-11-1267.
56. A. La Porta, and R. Slusher, "Squeezing limits at high parametric gains", Phys. Rev. A **44**, 2013–2022(1991), http://link.aps.org/doi/10.1103/PhysRevA.44.2013.
57. A. I. Lvovsky, "Continuous-variable optical quantum-state tomography", Rev. Mod. Phys. 81, 299–332(2009), http://link.aps.org/doi/10.1103/RevModPhys.81.299.
58. H. Hansen, T. Aichele, C. Hettich, P. Lodahl, A. I. Lvovsky, J. Mlynek, and S. Schiller, "Ultrasensitive pulsed, balanced homodyne detector: application to time-domain quantum measurements", Opt. Lett. **26**, 1714–1716(2001), http://www.opticsinfobase.org/abstract.cfm?URI=ol-26-21-1714.
59. J. P. Gordon, W. Louisell, and L. Walker, "Quantum Fluctuations and Noise in Parametric Processes. II", Phys. Rev. Lett. **129**, 481–485(1963), http://link.aps.org/doi/10.1103/PhysRev.129.481.
60. W. Wagner, and R. Hellwarth, "Quantum Noise in a Parametric Amplifier with Lossy Modes", Phys. Rev. A **133**, 915–920(1964), http://link.aps.org/doi/10.1103/PhysRevA.133.915
61. R. Byer, and S. Harris, "Power and Bandwidth of Spontaneous Parametric Emission", Physical Review **168**, 1064–1068(1968),http://link.aps.org/doi/10.1103/PhysRev.168.1064.
62. T. Hirano, K. Kotani, T. Ishibashi, S. Okude, T. Kuwamoto, "3 dB squeezing by single-pass parametric amplification in a periodically poled $KTiOPO_4$ crystal", Opt. Lett. **30**, 1722–1724(2005), http://www.opticsinfobase.org/abstract.cfm?URI=OL-30-13-1722.
63. C. Kim, R. D. Li, and P. Kumar, "Deamplification response of a traveling-wave phase-sensitive optical parametric amplifier", Opt. Lett. **19**, 132–134(1994), http://www.opticsinfobase.org/abstract.cfm?URI=ol-19-2-132.
64. J. Appel, D. Hoffman, E. Figueroa, and A. I. Lvovsky, "Electronic noise in optical homodyne tomography", Phys. Rev. A **75**, 035802(2007), http://link.aps.org/doi/10.1103/PhysRevA.75.035802.
65. R. Tualle-Brouri, A. Ourjoumtsev, A. Dantan, P. Grangier, M. Wubs, and A. Sørensen, "Multimode model for projective photon-counting measurements", Phys. Rev. A **80**, 013806(2009), http://link.aps.org/doi/10.1103/PhysRevA.80.013806.
66. R. Demkowicz-Dobrzanski, U. Dorner, B. J. Smith, J. S. Lundeen, W. Wasilewski, K. Banaszek, I. A. Walmsley, "Quantum phase estimation with lossy interferometers", Phys. Rev. A **80**, 013825(2009), http://link.aps.org/doi/10.1103/PhysRevA.80.013825.
67. Z. Hradil, "Quantum-state estimation", Phys. Rev. A **55**, R1561–1564(1997), http://link.aps.org/doi/10.1103/PhysRevA.55.R1561.
68. M. Ježek, J. Fiurášek, and Z. Hradil, "Quantum inference of states and processes", Phys. Rev. A **68**, 012305 (2003), http://link.aps.org/doi/10.1103/PhysRevA.68.012305.
69. A. I. Lvovsky, "Iterative maximum-likelihood reconstruction in quantum homodyne tomography", Journal of Optics B **6**, 556–559 (2004), http://iopscience.iop.org/1464-4266/6/6/014/.
70. P. Marek, and M. S. Kim, "Suitability of the approximate superposition of squeezed coherent states for various quantum protocols", Phys. Rev. A **78**, 022309(2008), http://link.aps.org/doi/10.1103/PhysRevA.78.022309.

## 1. Introduction

Quantum information processing solely based on Gaussian states and Gaussian operations is a largely matured field of research. The preparation of squeezed states - the ubiquitous resource in many Gaussian protocols - has experienced large progress in recent years. States with a high purity or a high degree of squeezing have been produced [1, 2, 3, 4]. Moreover, Gaussian projectors can be implemented using homodyne detection, which is capable of reaching near-unity detection efficiency [1]. Finally, Gaussian displacement operations combined with low-noise linear feedback have been implemented with high quality [5, 6, 7, 8]. This progress has lead to the implementation of various Gaussian protocols such as quantum teleportation [9], quantum key distribution [10], quantum cloning [11], quantum secret sharing [12] and quantum computation [13, 14].

However, several no-go theorems exist for systems consisting of purely Gaussian states and Gaussian operations. With this constrained set of states and operations it is impossible to perform entanglement distillation [15, 16, 17], quantum error correction [18], universal quantum computing [19, 20], quantum bit commitment [21], and to violate Bell's inequality [22]. To realize these protocols, non-Gaussian approaches are required. This non-Gaussianity can be injected into the system at different stages. It can enter through a non-Gaussian measurement strategy [23, 24, 25], non-Gaussian noise characteristics [26], or it can be incorporated through a non-Gaussian state preparation strategy [27, 28, 29, 30, 31, 32].

Important examples of a pure non-Gaussian state are the photon number eigenstates, the Fock states $|n\rangle, (n=1, 2...)$. Such states have been prepared and fully characterized in optical systems using SPDC followed by a non-Gaussian heralding measurement [33, 34, 35, 36, 37, 38]. Another family of non-Gaussian states, which has gained much interest in recent years, are the Schrödinger cat states which are superpositions of two coherent states of different phase, $|\alpha\rangle \pm |-\alpha\rangle$. Despite the constituents being Gaussian, the superposition exhibits strong non-Gaussianity which is sufficient for the realization of various protocols, examples being the realization of quantum computation[27, 39, 40, 41], error correction of Gaussian noise[30] and the violation of Bell's inequality [31].

It has been demonstrated that such coherent state superpositions (CSS) with a moderate amplitude $\alpha \lesssim 1$ can be well approximated by a photon-subtracted squeezed state [42], or equivalently, a squeezed single-photon state. Fidelities between the ideal CSS state and the photon-subtracted squeezed state as a function of the excitation, $\alpha$, of the CSS and the degree of squeezing of the squeezed state are shown in Fig. 1. The moderate amplitude of the CSS can be non-deterministically amplified to a larger amplitude CSS by means of linear interference and heralding based on photon counting or homodyne detection events [28, 43, 44, 45]. A large CSS can also be prepared through a conditional homodyne measurement on a Fock state in which the amplitude of the CSS state scales with the number of photons in the Fock state [46].

Inspired by these ideas for the generation of CSS and motivated by the potential applications, various groups have realized photon subtraction with squeezed states. These implementations have been carried out either with continuous-wave (CW) or pulsed light sources, and with wavelengths ranging from the near-infrared to the telecommunication regime. The photon subtraction has been carried out using an asymmetric beamsplitter that reflects a small portion of the light in which a photon is measured and thus subtracted from the squeezed state. The measurement has been realized with single-photon avalanche photodiodes (APD) as well as with photon-number-resolving transition edge sensors [35, 46, 47, 48, 49, 50, 51, 52].

The largest directly measured value of the Wigner function negativity is $-0.171$ and it is obtained using continuous wave (CW) squeezed states [53]. Using pulsed instead of CW squeezed light, the reported negativities as well as the purities of the generated non-Gaussian states are much lower. However, despite the lower quality of the generated states, there has been much interest in pulsed experiments due to the relative simplicity of the experimental setup and the inherent temporal confinement of the generated states. Previous pulsed experiments on generating CSS have employed femtosecond pulsed lasers and very short (sub-millimeter long) nonlinear crystals [35, 50].

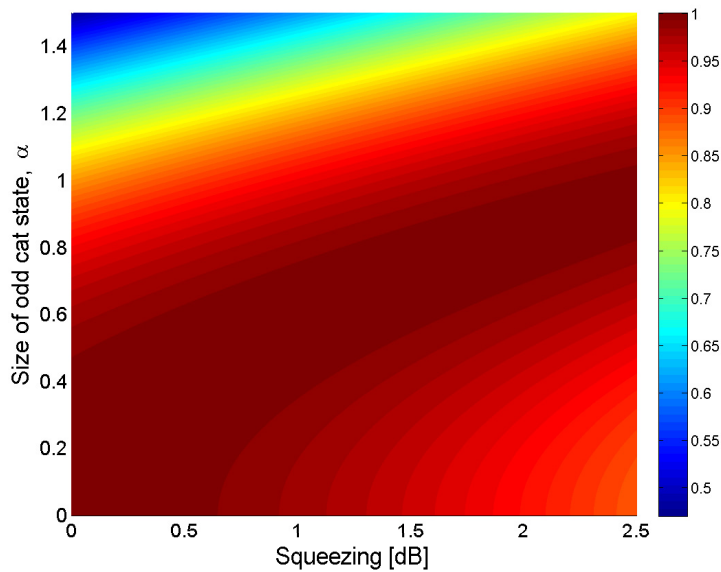

Fig. 1. Fidelity between a photon-subtracted squeezed state and an ideal coherent state superposition, $|\alpha\rangle - |-\alpha\rangle$ for different degrees of squeezing. It can be seen that the fidelity remains high, $F > 0.9$, for $\alpha$ up to 1, provided the squeezing degree is not too large.

Here we present experiments on CSS generation using picosecond laser pulses and a 3-mm long quasi-phase matched PPKTP crystals for squeezed light production. By using picosecond instead of femtosecond pulses, the group velocity dispersion is small and thus a longer crystal for squeezed light production could be used.

## 2. Experiment

A CSS state with a small amplitude can be approximated by a photon subtracted squeezed state, and it can be fully characterized by means of its Wigner function which is obtainable by homodyne tomography. In the following we present the different parts of our experimental setup to generate and characterize a CSS. We introduce the two required parametric processes (up-conversion and down-conversion), the photon subtraction setup, and finally the homodyne detector. We also briefly discuss a simple model for predicting the performance of the experiment.

The experimental setup is shown in Fig. 2. We used a cavity-dumped Titanium-Sapphire pulsed laser, which produced nearly Fourier-transform-limited pulses with duration 4.6 ps at 830 nm with an average energy up to 40 nJ and a repetition rate of 815 kHz. The spectral width of the pulses was measured to be 0.16 nm, corresponding to a spectral width of 70 GHz with a center wavelength of 829.7 nm. A fraction of about 10% was used as a local oscillator (LO) for homodyne detection, a weak seed beam was directed to the parametric down-conversion process for alignment and the remaining

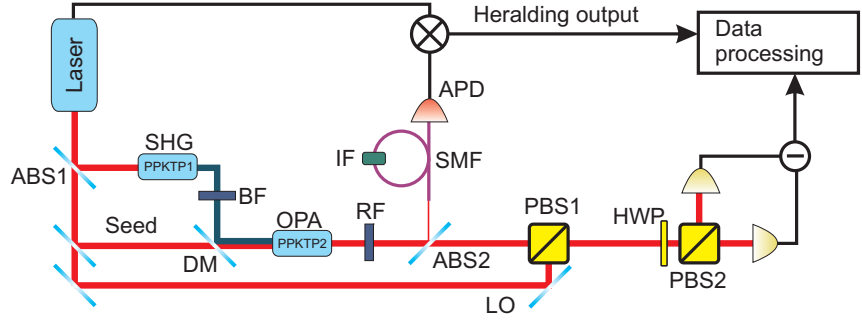

Fig. 2. Schematic of the experimental setup. The cavity dumped laser emits 4.6 ps optical pulses at 830 nm with a repetition frequency of 815 kHz. ABS1: 90/10 beamsplitter, BF: blue filter, DM: dichroic mirror, RF: red filter, ABS2: Asymmetric beamsplitter with $R = 7.7\%$, SMF: single-mode fiber, IF: interference filter, APD: avalanche photodiode, PBS: polarizing beamsplitter, HWP: half-wave plate.

part was directed to a frequency doubling process.

## 2.1. Frequency doubling

For frequency doubling a 3-mm long periodically poled KTP crystal (PPKTP1) was used. The crystal poling-period was chosen for 1st order quasi phase-matching corresponding to a poling period of $\Lambda \sim 3.8$ μm with all fields polarized along the crystal z-axis. The length of the crystal was chosen to be 3mm which was a compromise between having a large interaction length and avoiding phase mismatch as a result of group velocity dispersion.

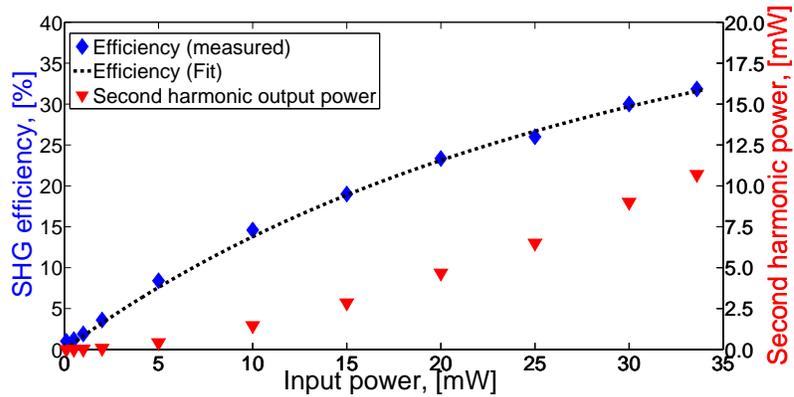

Fig. 3. (a) Blue diamonds correspond to the second-harmonic generation conversion efficiency for different input powers of the fundamental beam, $P_F$. These points are fitted with $\eta = \eta_\infty \tanh^2(g\sqrt{P_F})$ with $\eta_\infty = 53\%$ and $g = 0.18$. Red triangles are the total generated SHG power for different power levels of the fundamental beam.

The beam waist was set to $w_0 \sim 90$ μm, thus achieving a weak focusing condition with the depth-of-focus ($2z_0 \sim 60$ mm) being 20 times longer than the crystal length [54, 55]. The second harmonic conversion efficiency, $\eta$ was investigated as a function of input power of the the fundamental beam and the result is displayed in Fig. 3. A maximum frequency doubling efficiency of 32 % was achieved for an incident average power of 33 mW. The spectral properties of the resulting frequency doubled light was investigated by a wavelength meter and it was measured to have a center wavelength of 414.8 nm and a bandwidth of $\sim$0.1 nm, corresponding to a spectral width of about 174 GHz.

## 2.2. Parametric down-conversion

After the frequency doubling crystal, the residual light at 830 nm was filtered out using a series of filters. The filtered blue light was then focused into a second PP-KTP crystal (PPKTP2) used to generate squeezed vacuum through the SPDC. It has been shown that for single pass pulsed SPDC experiments the gain-product $g_{min}g_{max}$, where $g_{min}(g_{max})$ is the attenuation(amplification) factor is enhanced by defocussing the pump [55, 56]. This was confirmed in our setup, and the waist of the pump was set to $w_{0,p} \sim 150$ µm and a depth-of-focus of $2z_0 \sim 340$ mm. The pump can thus be regarded as a plane wave within the length of the crystal, leading to an improvement of the degree of squeezing [56]. After the SPDC, the remaining pump was filtered out using a series of filters. The generated squeezed vacuum was directed to an asymmetric beamsplitter (ABS2) with a reflectivity of $R \approx 7.7\%$. The reflected part was directed to an avalanche photodiode, while the transmitted part was subjected to full quantum state tomography by means of time-domain balanced homodyne detection (TD-BHD) [57, 58].

In pulsed experiments, squeezing is often generated in a single pass configuration without the use of enhancement cavities. Thus the squeezing is generated in many different spatial and temporal modes [59, 60, 61]. The mode (and thus the degree of squeezing) being measured by the homodyne detector depends on the spatio-temporal profile of the local oscillator: the mode of the squeezing spectrum that spatially and temporally overlaps with the LO will be measured by the homodyne detector. The amount of measured squeezing can be optimized by injecting a weak seed beam, corresponding to the mode of the LO, into the SPDC crystal and studying the classical parametric (de-)amplification of the seed beam. Depending on the relative phase between the seed and the pump, the seed can be either amplified or de-amplified. As described above the beam waist of the pump in the PPKTP was set to $w_{0,p} = 150$ µm. Optimal phase-matching between the two waves is obtained for $w_{0,p}/w_{0,s} = \sqrt{2}$ [55], and thus we set the beam waist of the seed ($w_{0,s} = 106$ µm). We achieved an optimal de-amplification of $g_{\min} = 0.37$ and an amplification of $g_{\max} = 4.6$ for a pump power of 9 mW. A full characterization of the gains, $g_{\{\min,\max\}}$, as a function of pump power is shown in Fig. 4.

Using a simple model we find [62],

$$g_{\{min,max\}} = \epsilon \exp(\{+2r, -2r\}) + (1-\epsilon) \quad (1)$$

where $\exp(\pm 2r)$ is the intrinsic gain and $\epsilon$ is a parameter describing the spatial overlap between the seed and the pump. $\epsilon$ was experimentally estimated to be $0.77 \pm 0.01$ (see below). Through power-shape fitting, the dependence of $r$ on the the pump power is found to be $r = 0.28\sqrt{P_p}$. This measurement of the classical gain also shows that within the gain range of our experiment, the effect of gain-induced diffraction is negligible [56, 63].

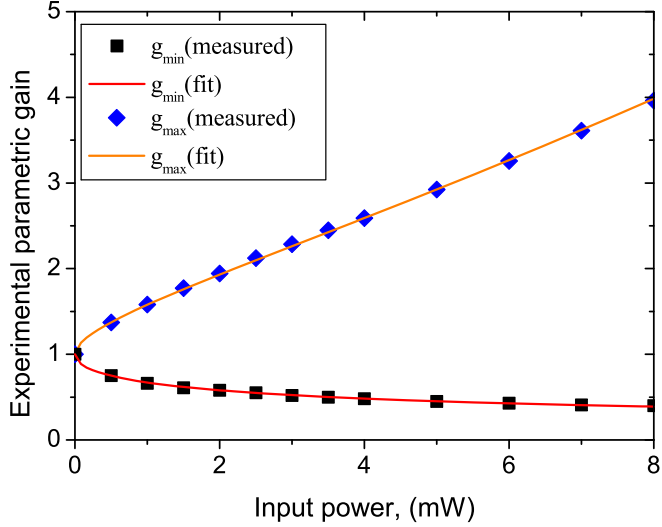

Fig. 4. Classical parametric gain versus average pump power, $P_p$. The blue squares correspond to the measured deamplification values whereas the blue diamonds are the measured amplification values. The red and green dotted lines are fits using Eq. (1), with the intrinsic parametric gain $r = 0.28\sqrt{P_p}$ and $\epsilon = 0.77 \pm 0.01$.

## 2.3. Time-Domain balanced homodyne detection

The measurement setup for homodyne detection is shown in Fig. 2. The transmitted squeezed vacuum state was superimposed with the LO at PBS1, the combined optical pulses were split at PBS2 and the resulting two beams were focused onto a pair of PIN photodiodes (Hamamatsu, S3883, quantum efficiency of $\eta_{ph} = 95 \pm 2$ %). Using a HWP, the splitting ratio of PBS2 was tuned to balance the homodyne detector. The detector used in the experiment is based on the design by Hansen et al. [58]. Its output was recorded by a digital oscilloscope (LeCroy, LT374L) using the cavity dumper signal from the laser as a trigger. The final quadrature value is then extracted by integrating the signal over the individual pulses.

The integration requires a well defined pulse window, $T_W$, which is determined by the repetition rate of the laser $f_{Rep} = 815$ kHz yielding $T_W \approx 1.2$ µs. The detector has a bandwidth of 2 MHz which is confirmed by the generation of 500 ns wide electronic pulses resulting from the detection of the picosecond optical pulses. Since electronic pulse is shorter than the pulse window, only a fraction of the pulse window contains valuable information. As a result, only a part of the measured pulse contributes to the integration, used to extract the quadrature values. We investigated the signal-to-noise ratio (SNR) (shot-noise variance to electronic noise variance) of the detection scheme for various choices of a weight function folded with the measured pulse. It was found that a simple boxcar-average, encompassing about 40 % of the measurement window, was an optimal choice. The shot noise reference is obtained by measuring a vacuum input state. The reference level is known to increase linearly with the local oscillator (LO) power. To verify that the system was indeed shot-noise limited, we measured the shot-noise as a function of the LO power, see Fig. 5(a). The electronic noise was measured to be 3.7 mV$^2$ which corresponds to 530 electrons/pulse. In Fig. 5(a) it is clearly seen that the shot-noise depends linearly on the LO power, and the gain of the detector was found to be 13.6 mV$^2$/10$^6$ photons per pulse. In Fig. 5(b) the ratio between the shot noise and the electronic noise (electronic noise clearance) is plotted as a function of the LO power. It can be seen that the noise clearance surpasses 23 dB when the LO pulse contains more than $70 \times 10^6$ photons (corresponding to a power of $\sim 12\mu$W). This corresponds to an electronic noise equivalent quantum efficiency of $\eta_{el} \geq 99.5 \pm 0.5\%$ [64].

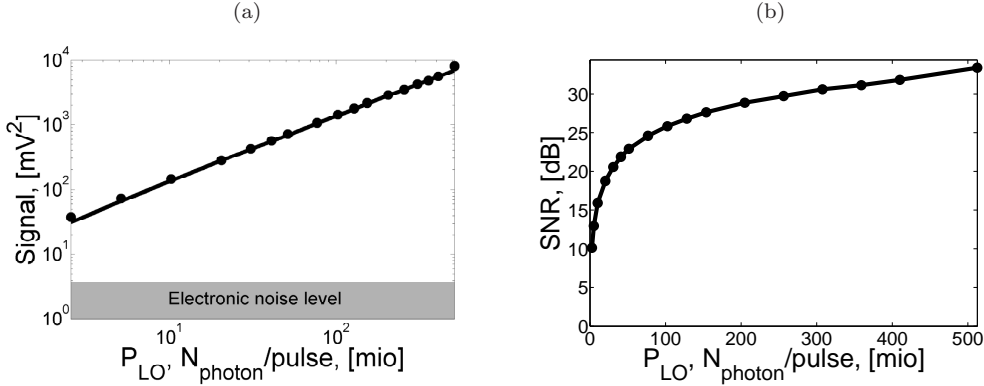

Fig. 5. (a) The quadrature variance measurement of the shot noise as a function of the LO power. (b) the signal-to-noise ratio (SNR) of the shot noise variance as a function of the LO power.

The overall homodyne detection efficiency $\eta_{hd}$ is given by

$$\eta_{hd} = \eta_{op}\eta_{mm}^2\eta_{ph}\eta_{el} \qquad (2)$$

where $\eta_{op}$ is the propagation efficiency of the state through optical components and $\eta_{mm}$ is the degree of mode-matching between the LO and the squeezed pulse. These values were measured to be $\eta_{op} = 0.90 \pm 0.02$ and $\eta_{mm} = 0.95 \pm 0.02$, giving a total homodyne efficiency of $\eta_{hd} = 0.77 \pm 0.02$.

*2.4. Photon subtraction*

The reflected photons from ABS2 were detected by a fiber coupled APD (Perkin-Elmer SPCM-AQR-14-FC). Using this signal as a trigger we conditionally prepared a photon subtracted squeezed state. To reduce the effect of detector dark counts, the trigger signal for the homodyne measurement is derived by correlating the APD signal with the cavity dumper signal (see Fig. 2). By setting the coincidence window to 120 ns - corresponding to 1/10 of the total measurement window - we achieved a 10-fold decrease of the detector dark counts, resulting in a dark count rate of $2.0 \pm 0.5$ $s^{-1}$. To ensure that the APD detection events are spatially and spectrally matched with the optical mode of the LO it is necessary to employ filtering in the APD arm. The spectral filtering was carried out using a Fiber-coupled Tunable Fabry-Perot (FP) Filter Cavity (Micron Optics, FFP-TF-830-005). It was coupled via two 1.5-meter-long single mode fiber pigtails for spatial filtering. The bandwidth of the filter was 22 GHz corresponding to 0.05 nm, and the central wavelength could be tuned using a voltage supply. The total detection efficiency of the heralding channel was estimated to be about $10 \pm 5$ % including the coupling efficiency to the fiber, the peak transmission of the FP filter and the detection efficiency of the APD. The total gated signal trigger count rate varied from 400 $s^{-1}$ up to 4000 $s^{-1}$ for pump powers of 1 mW up to 8 mW.

*2.5. Gaussian model for estimation of photon-subtracted squeezed state*

In order to predict the performance of the photon-subtraction experiment, we derived a simple model Refs. [35, 48, 65]. The model takes into account various experimental imperfections, which could compromise the quality of the prepared output states. The analysis is broken into three parts. The first part is the generation of the squeezed state. The second is the tap-off on the asymmetric beamsplitter, and projection onto the on-off click detector with a filter in front. Finally, the third part is the imperfect homodyne detector used for characterization. We choose to work with the Wigner quasi-probability distributions since they provide a convenient framework for such types of models. In the Wigner picture the vacuum state is given by a simple Gaussian distribution in the quadrature variables, $\hat{X} = [x,p]^T$.

$$W_0(\hat{X}) = \frac{e^{-x^2-p^2}}{\pi} \qquad (3)$$

A squeezed state can be written in the same way with the variables rescaled according to the quadrature variances, $V_{x,p}/2$:

$$W_s(\hat{X}) = \frac{e^{-\frac{x^2}{V_x}-\frac{p^2}{V_p}}}{\pi\sqrt{V_x V_p}} \qquad (4)$$

where the Heisenberg uncertainty principle constrains the variances as $V_x V_p \geq 1$. The squeezed state is split on an asymmetric beamsplitter with reflectivity $R$, and one part is measured using the positive operator value measure (POVM) element, $\hat{\Lambda} = \hat{1} - |0\rangle\langle 0|$,

$$W_{out}(\hat{X}_1) = 2\pi \int W(\hat{X}_1, \hat{X}_2) W_\Lambda(\hat{X}_2) \, d\hat{X}_2 \qquad (5)$$

where $W_\Lambda(\hat{X}_2)$ is the Wigner function for the POVM element and $W(\hat{X}_1, \hat{X}_2)$ is the Wigner function of the state after the beamsplitter. In Sec. 2.4 we described how proper filtering in the heralding arm was necessary in order to ensure that identical modes were detected by the APD and the homodyne detector. However, in practice, some false modes will be detected by the APD. This will be modeled by the modal purity parameter $\Xi$. It describes the probability that the photon detected from the APD came from the targeted optical mode. The output from the system can then be expressed as follows,

$$W_{out,\Xi}(\hat{X}) = \Xi W_{out}(\hat{X}) + (1-\Xi) W_s(\hat{X}) \qquad (6)$$

where $W_s(x,p)$ is the state we see when the APD detections are uncorrelated with the optical mode used in the experiment.

**3. Experimental results**

*3.1. Squeezed vacuum*

The squeezed vacuum produced by the SPDC was characterized by using a digital oscilloscope with which we acquired ∼ 65200 quadrature values. Using a sample rate of 100 MS/s for 80 ms we acquired 120 pts. per pulse. The cavity dumper signal (rate of 815 kHz) was used to set the time window for each pulse acquisition, see Sec. 2.3. The relative phase between the LO and the quantum state was scanned by a saw-tooth modulation applied to a piezocrystal attached to a highly reflecting mirror placed in the LO arm. The minimum and maximum quadrature variances of the squeezed pulses were measured as a function of pump power and the measured values are shown in Fig. 6.

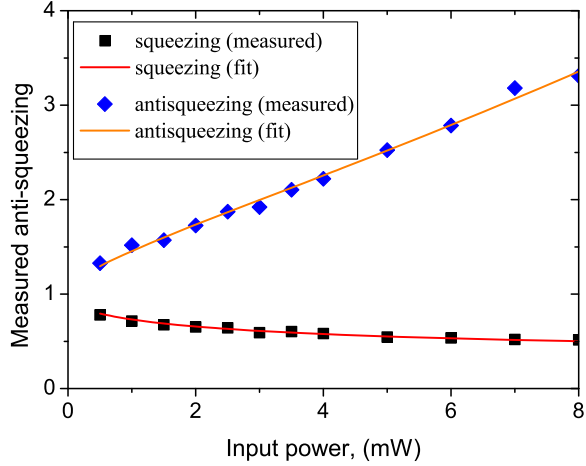

Fig. 6. The measured squeezing and anti-squeezing as a function of the average pump power, $P_p$. The blue diamonds are the measured maximum variances (anti-squeezing), the blue squares are the measured minimum variances (squeezing), and the green aned red dotted lines are fittings to Eq. (1) where $r = 0.28\sqrt{P_p}$ and $\eta = 0.62 \pm 0.01$.

Assuming a loss model, the squeezing and anti-squeezing variances, $V_{\{min,max\}}$, can be fitted to a simple relation,

$$V_{\{min,max\}} = (\eta(g_{min}, g_{max}) + (1-\eta))V_{vac} \qquad (7)$$

where $g_{\{min,max\}}$ is the parametric gain, $\eta$ is the total detection efficiency, and $V_{vac}$ is the quadrature variance of the vacuum state. By setting the parametric gain to the values found in Sec. 2.2, the expressions can be fitted to the experimental data using the efficiency $\eta$ as the fitting parameter. We find reasonable fits to both series of data for $\eta = 0.62$. This value is lower than the experimentally accessed one of 77%, see Sec. 2.3. Part of the discrepancy is caused by the tap-off beamsplitter which adds 8% loss to the squeezed states (not accounted for in section 2.3). The remaining discrepancy of 9%, we speculate to be resulting from a mismatch between the temporal modes of the LO and the squeezed vacuum [54, 66]. However, as this loss effect has not be carefully studied and localized, we will be using $\eta = 0.77$ as the estimated detection loss in the following sections. This value will be used to correct the experimental data for losses.

Using the experimental parameters in Fig. 6 as well as the formalism given in Sec. 2.5, the expected properties of the photon-subtraction squeezed state is theoretically predicted. We calculate the fidelity of the CSS state, $F_{odd}$, its amplitude, $\alpha$, and the negativity of its Wigner function at the origin, $W(0,0)$, at four pump power levels. The results are listed in Table 1 and Table 2 (corrected for losses). It is clear that any one of the photon subtracted squeezed states are expected to exhibit strong non-Gaussianity with relatively large negativities of the Wigner functions.

| $P_p$ [mW] | $F_{odd}$ | $\alpha$ | $W(0,0)$ |
|---|---|---|---|
| 2.0 | 0.64 | 0.87 | -0.09 |
| 4.0 | 0.58 | 1.05 | -0.06 |
| 6.0 | 0.55 | 1.20 | -0.04 |
| 8.0 | 0.52 | 1.32 | -0.03 |

Table 1. Estimation of the parameters characterizing the photon subtracted squeezed state for different pump powers. The predictions are based on the measured values for the squeezed vacuum states corrected for losses associated with the detection ($\eta = 0.77$) as well as the asymmetric beamsplitter (8% loss). In these estimates we have set the mode match parameter $\Xi$ to unity for all power levels.

| $P_p$ [mW] | $F_{odd}$ | $\alpha$ | $W(0,0)$ |
|---|---|---|---|
| 2.0 | 0.84(0.99) | 0.99(1.06) | -0.21(-0.32) |
| 4.0 | 0.74(0.89) | 1.18(1.27) | -0.16(-0.26) |
| 6.0 | 0.69(0.83) | 1.34(1.44) | -0.13(-0.23) |
| 8.0 | 0.63(0.76) | 1.45(1.56) | -0.11(-0.20) |

Table 2. Estimation of the parameters characterizing the photon subtracted squeezed state for different pump powers and after correcting for the homodyne losses of 0.77. For comparison, the numbers corrected for 0.62/0.92 efficiency are also given in brackets.

### 3.2. Photon-subtracted squeezed vacuum

Next, we prepared photon-subtracted squeezed states for different pump powers ranging from 2 to 8 mW. In this range, the photon detection rate of the APD varied from 400 s$^{-1}$ to 4000 s$^{-1}$. Each time a detection event from the APD was correlated with the sync signal from the laser, a trigger signal was derived and sent to the oscilloscope, see Sec. 2.1. For every trigger signal, the digital oscilloscope sampled the homodyne signal for 1 μs with a sampling rate of 250 MS/s, making up a single measurement segment. The segments were stored consecutively until 4000 data segments filled up the memory of the oscilloscope. The quadratures were extracted in the same way as for the squeezing measurement. During one measurement series the relative phase between the LO and the quantum state was scanned over a range of $0-3\pi$.

We used maximum likelihood estimation to reconstruct the prepared quantum state [67, 68, 69]. Before the reconstruction, the acquired quadrature files were concatenated, and the entire batch of quadratures measurements was used for reconstruction. In order to reconstruct the quantum state, an estimation of the phase reference was required. Since the phase was scanned, we did not have a stable phase reference. In order to extract the phase information the quadrature data was stored in bins of 100 quadratures and the variance of each bin was evaluated. The phase of bin $i$ can be estimated by comparing its variance to the minimum and maximum variances using the relation $V_i(\theta_i) = V_{min}\cos^2\theta_i + V_{max}\sin^2\theta_i$, where $\theta_i$ is the phase associated with bin $i$. From this expression we obtained an estimate for the phases in each bin. Without loss of generality, the assigned phase was chosen between 0 and $\pi/2$. Having assigned a phase to the quadratures, we performed maximum likelihood reconstruction. We reconstructed density matrices and from those we calculated the Wigner quasi-probability distributions, see Fig. 7 (without loss-correction) and Fig. 8 (corrected for losses).

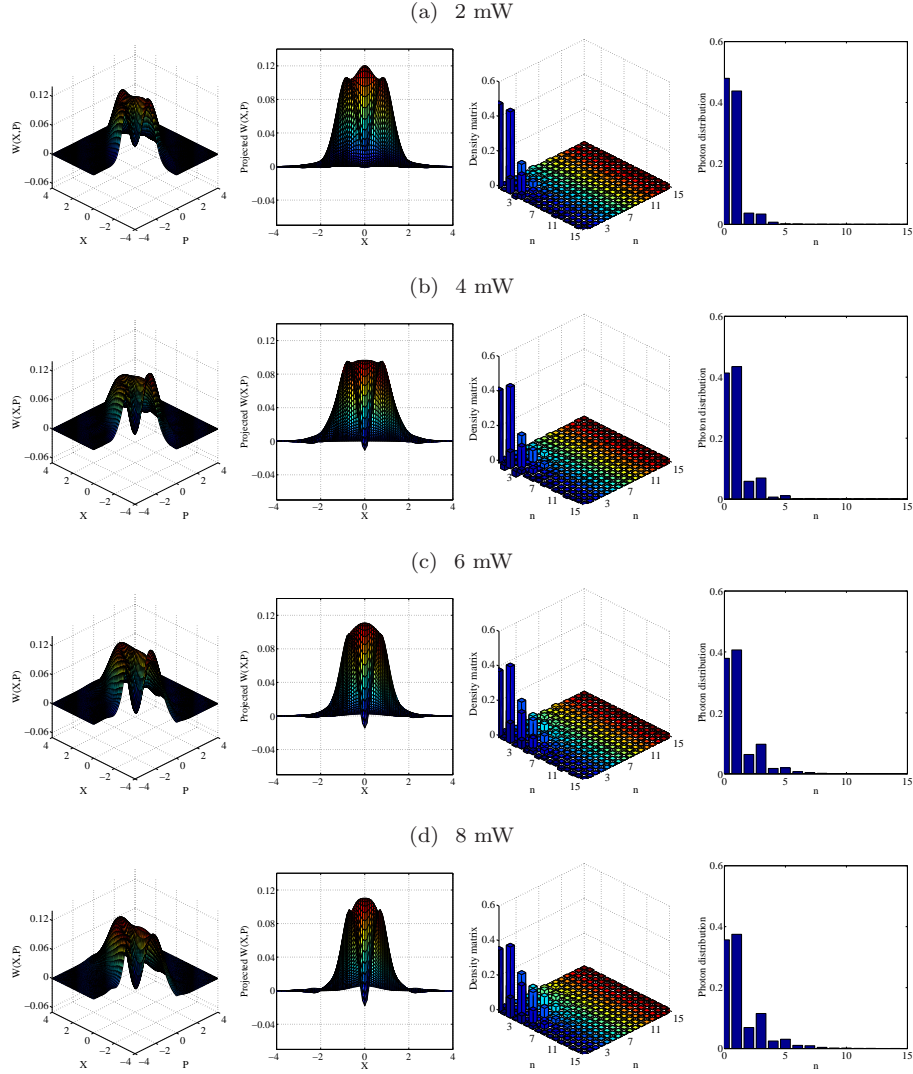

Fig. 7. Plot of the Wigner functions, the projected Wigner functions, the density matrices and the photon distributions of the reconstructed states for pump powers from 2 mW to 8 mW. There are no corrections for losses in these plots.

From Fig.7 we see that the generated states are non-Gaussian and non-classical. Moreover, it is evident that the Wigner functions become more squeezed as the pump power increases while the dip around the origin retains its structure. The dip attains a negative value for all realizations if the measurement results are corrected for losses. The maximum measured negativity for the uncorrected data is $W(0,0) = -0.023$ which corresponds to a negativity of $W(0,0) = -0.063$ after loss correction. The fidelities between the experimentally produced states and the ideal cat states (maximized over the excitations $\alpha$) are summarized in Table 3.

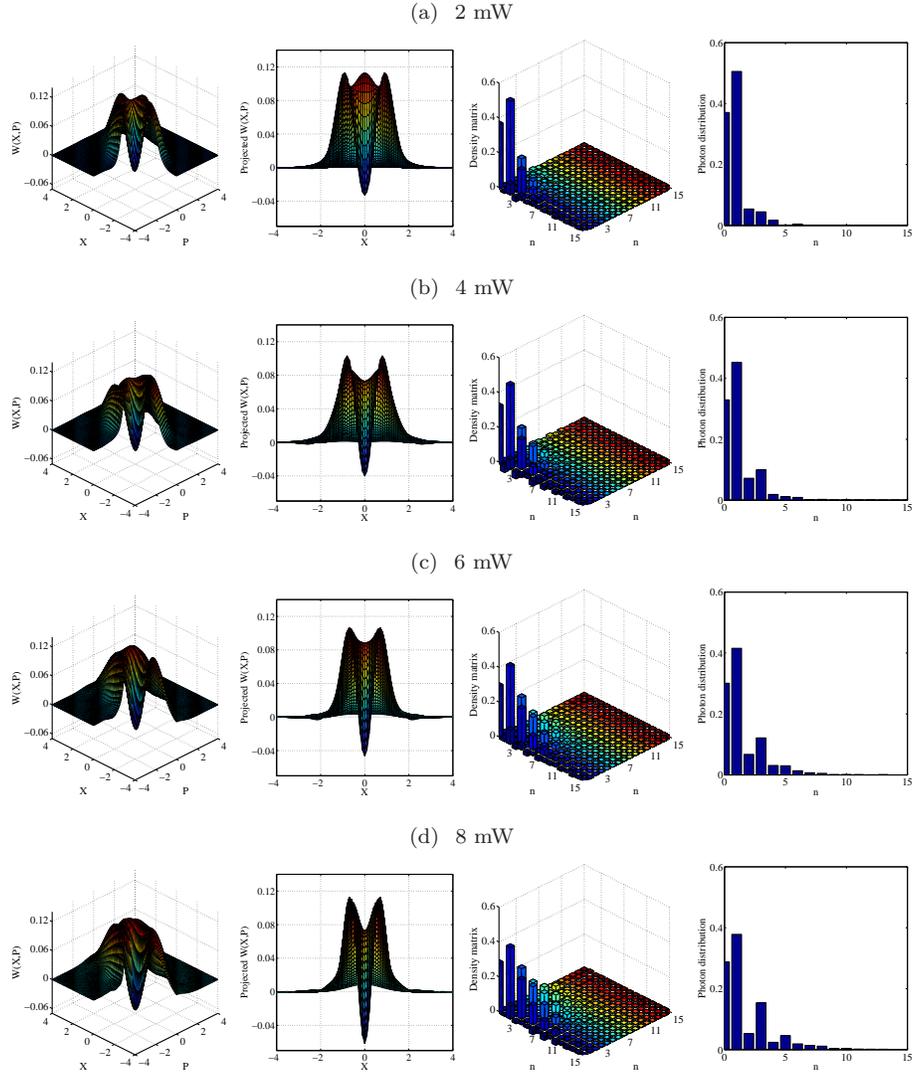

Fig. 8. Plot of the Wigner functions, the projected Wigner functions, the density matrices and the photon distributions of the loss-corrected reconstructed states for pump powers from 2 mW to 8 mW.

By comparing the results in Table 3 with the predictions in Table 1 we see that the fidelities, as well as the negativities, are generally smaller than predicted. The discrepancy, however, gets smaller at higher pump powers. This effect is caused by slow instabilities of the experimental setup which become significant for longer measurement runs as is the case for low pump powers where the run time is about 10s (to acquire 4000 quadrature values). For high powers, however, the measurement time (1-2s) is shorter and thus the influence of instabilities is less pronounced. The main source of instability is a mechanical drift of the filtering Fabry-Perot cavity which was not actively stabilized during the measurement. A drift of the cavity results in the detection by the APD of the frequency modes, which are different from the ones measured at the homodyne detector, which results in degradation of the performance. This corresponds to a lower value of the parameter $\Xi$ in Eq. (6). To estimate values for the mode match parameter $\Xi$ for the different power levels, we fit the theoretical predictions to the actual measurement results by using $\Xi$ as a fitting parameter. The obtained values of $\Xi$ for which the theoretical fidelities and Wigner function negativities match the experimental ones are shown in Table 3, which shows that the mode matching parameter is increasing for increasing pump powers.

| $P_p$ [mW] | $F_{odd}$ | $\alpha$ | $W(0,0)$ | $\Xi$ |
|---|---|---|---|---|
| 2.0 | 0.46 | 0.77 | 0.016 | 0.72 |
| 4.0 | 0.47 | 0.88 | $-0.011$ | 0.86 |
| 6.0 | 0.49 | 1.08 | $-0.018$ | 0.91 |
| 8.0 | 0.49 | 1.19 | $-0.023$ | 0.96 |

Table 3. Parameters characterizing the prepared photon subtracted squeezed states.

Incorporating the generalized Bernoulli transformation into the maximum likelihood algorithm, the homodyne detection inefficiency can be corrected [69]. As mentioned above, we used the conservative estimate of the detection efficiency of 77% for the correction in order to avoid overestimating the negativities of the corrected Wigner functions. The Wigner functions after correction are displayed in Fig. 8, and the results for the fidelities, negativities and sizes are summarized in Table 4.

| $P_p$ [mW] | $F_{odd}$ | $\alpha$ | $W(0,0)$ |
|---|---|---|---|
| 2.0 | 0.55 | 0.83 | $-0.033$ |
| 4.0 | 0.53 | 1.01 | $-0.042$ |
| 6.0 | 0.54 | 1.17 | $-0.050$ |
| 8.0 | 0.56 | 1.32 | $-0.063$ |

Table 4. Parameters characterizing the prepared photon subtracted squeezed states after correction for imperfect detection.

## 4. Conclusion

We have presented a method of the preparation of photon-subtracted squeezed states in a system based on picosecond pulsed laser pulses. It is based on generating squeezed vacuum from SPDC in a PPKTP crystal followed by single photon subtraction, enabled by the reflection of a single photon on an asymmetric beamsplitter and its detection by the APD. Various states were produced with varying degree of squeezing.

The resulting states were fully characterized by homodyne tomography with which the Wigner functions and density matrices were reconstructed. We found a maximum negativity of $W(0,0) = -0.023$ without any loss-corrections and $W(0,0) = -0.063$ after loss-correction. The negativity appeared to be largest for the largest degrees of squeezing. It is attributed to the shorter measurement time associated with larger squeezing and thus greater robustness to instabilities of the setup.

Using a simple theoretical model, we performed a comparison between theoretical and experimental results. A discrepancy is attributed to a mismatch between the modes measured by the APD and the homodyne detector. By fitting the experimental results, we estimated a value for the mismatch. It was found to depend on the degree of squeezing and therefore on the measurement time: the mismatch was found to become increasingly severe as the degree of squeezing was lowered corresponding to a longer measurement time. As mentioned above, the extended measurement time resulted in instabilities during the measurement time, thus degrading the performance. In particular, the temporal filter prior to single photon detection was drifting thereby resulting in the detection of different temporal modes during the acquisition time. In order to improve the performance of the setup, the temporal filter cavity must be actively stabilized.